\documentclass[final,5p,times,twocolumn]{elsarticle}
\usepackage{lineno,hyperref,amsmath,amsthm,amssymb,amsfonts,ragged2e,color,subfig}
\journal{``Plasma Physics Reports''}
\begin{document}
\begin{frontmatter}
\title{Electrostatic shock structures in a magnetized plasma having non-thermal particles}
\author{S. Jahan$^{*,1}$,  S. Banik$^{**,1,2}$, N.A. Chowdhury$^{***,3}$, A. Mannan$^{\dag,1}$, and A.A. Mamun$^{\S,1}$}
\address{$^1$Department of Physics, Jahangirnagar University, Savar, Dhaka-1342, Bangladesh\\
$^2$Health Physics Division, Atomic Energy Centre, Dhaka-1000, Bangladesh\\
$^3$Plasma Physics Division, Atomic Energy Centre, Dhaka-1000, Bangladesh\\
Email: $^{*}$jahan88phy@gmail.com, $^{**}$bsubrata.37@gmail.com, $^{***}$nurealam1743phy@gmail.com\\
$^{\dag}$abdulmannan@juniv.edu, $^{\S}$mamun\_phys@juniv.edu}
\begin{abstract}
A rigorous theoretical investigation has been made on the nonlinear propagation of dust-ion-acoustic shock waves in a multi-component magnetized pair-ion plasma having inertial warm positive and negative ions, inertialess non-thermal electrons and positrons, and static negatively charged massive dust grains. The Burgers' equation is derived by employing reductive perturbation method. The plasma model supports both positive and negative shock structures in the presence of static negatively charged massive dust grains. It is found that the steepness of both positive and negative shock profiles declines with the increase of ion kinematic viscosity without affecting the height, and the temperature of the electrons enhances the amplitude of the shock profile. It is also observed that the increase in oblique angle rises the height of the positive shock profile, and the height of the positive shock wave increases with the number density of positron. The application of the findings from present investigation are briefly discussed.
\end{abstract}
\end{frontmatter}
\section{Introduction}
\label{5:Introduction}
Positive ions are produced by the electron impact ionization while negative ions are produced
due to the attachment of electron with an atom \cite{Shukla2002}, and the existence of both
positive and negative ions or pair-ion (PI) can be observed in space plasmas, viz., cometary comae \cite{Chaizy1991},
upper regions of Titan's atmosphere \cite{Coates2007,Massey1976,Sabry2009}, plasmas in the D and F-regions of Earth's
ionosphere \cite{Massey1976,Sabry2009,Abdelwahed2016}) and also laboratory plasmas, viz., ($K^+$, $SF_6^-$) plasma \cite{Song1991,Sato1994}, ($Ar^+$, $F^-$) plasma \cite{Nakamura1984}, plasma processing reactors \cite{Gottscho1986}, plasma etching \cite{Sheehan1988}, combustion products \cite{Sheehan1988}, ($Xe^+$, $F^-$) plasma \cite{Ichiki2002}, neutral beam sources \cite{Bacal1979}, ($Ar^+$, $SF_6^-$) plasma \cite{Wong1975,Cooney1991,Nakamura1997},($Ar^+$, $O_2^-$) plasma, and Fullerene ($C_{60}^+$, $C_{60}^-$) plasma \cite{Oohara2003,Hatakeyama2005}, etc.
The dynamics of the plasma system and associated nonlinear electrostatic structures have been rigorously changed by the presence of massive
dust grains in the PI plasma (PIP) \cite{Shukla1992,Shukla2000,Baluku2010,Sultana2014}. Yasmin \textit{et al.} \cite{Yasmin2013}
studied the nonlinear propagation of dust-ion-acoustic (DIA) waves (DIAWs) in a multi-component plasma, and found that the shock profile associated with
DIAWs is significantly modified by the existence of dust grains. A number of authors also examined the effects of the positron to the
formation of solitary profile associated with electrostatic waves \cite{Rahman2015,Abdelsalam2008}. Rahman \textit{et al.} \cite{Rahman2015} studied the electrostatic solitary waves in electron-positron-ion plasma, and observed that the amplitude of the solitary profile increases with increasing the number density of positron. Abdelsalam \cite{Abdelsalam2008} investigated ion-acoustic (IA) solitary waves in a dense plasma, and demonstrated that the presence of the positron can cause to increase the amplitude of the solitary profile.

Cairns \textit{et al.} \cite{Cairns1995} first demonstrated the non-thermal distribution to investigate the effect of energetic particles
on the formation of IA shock profile, and introduced the parameter $\alpha$ in the non-thermal distribution for measuring the amount of deviation of non-thermal plasma species from Maxellian-Boltzmann distribution \cite{C1}. The non-thermal plasma species are regularly seen in the comtary  comae \cite{Chaizy1991}, Earth's ionosphere \cite{Massey1976}, and the upper region of the Titans \cite{Coates2007}, etc.
Haider \textit{et al.} \cite{Haider2019} investigated the IA solitary waves in the presence of
non-thermal particles, and observed that the width of the solitary profile decreases with increasing of ions non-thermality.
Pakzad and Javidan \cite{Pakzad2009} studied the dust-acoustic (DA) solitary and shock waves in a dusty plasma having non-thermal
ions, and reported that the amplitude of the wave increases with the decrease of the non-thermality of ions. Ghai \textit{et al.} \cite{Ghai2018} studied the DA solitary waves in the presence of non-thermal ions, and found that the height of the solitary wave decreases with the increase of $\alpha$.

Landau damping and the kinematic viscosity among the plasma species are the primary reason to the formation of shock profile associated with electrostatic waves \cite{Sabetar2015,Shahmansouri2014,Malik2020}. The existence of the external magnetic field is considered to be responsible to change the
configuration of the shock profile. Sabetkar and Dorranian \cite{Sabetar2015}  examined the effects of external magnetic field to the formation of the DA solitary waves in the presence of non-thermal plasma species, and found that the amplitude of solitary wave increases with the increase in the value of oblique angle. Shahmansouri and Mamun \cite{Shahmansouri2014} analysis the DA shock waves in a magnetized non-thermal dusty plasma and demonstrated that the amplitude of shock wave increases with increasing the oblique angle. Malik \textit{et al.} \cite{Malik2020} studied the small amplitude DA wave in magnetized plasma, and reported that the height of the shock wave enhances with oblique angle.
Bedi \textit{et al.} \cite{Bedi2010} studied DA solitary waves in a four-component magnetized dusty plasma, and highlighted that
both compressive and rarefactive solitons can exist in the presence of external magnetic field.
To the best knowledge of the authors, no attempt has been made to study the DIA shock waves (DIASHWs) in a magnetized PIP by considering kinematic viscosity of both inertial warm positive and negative ion species, and inertialess non-thermal electrons and positrons in the presence of static negatively charged dust grains. The aim of our present investigation is, therefore, to derive Burgers' equation and investigate DIASHWs in a magnetized PIP, and to observe the effects of various plasma parameters (e.g., mass, charge, temperature, kinematic viscosity, and obliqueness, etc.) on the configuration of DIASHWs.

The layout of the paper is as follows: The basic equations are displayed in section \ref{5sec:Governing equations}. The well known Burgers' equation has been derived in section \ref{5sec:Derivation of the Burgers equation}. Numerical analysis and discussion are presented in section \ref{5sec:Numerical Analysis and Discussion}. A brief conclusion is pinpointed in section \ref{5sec:Conclusion}.
\section{Governing equations}
\label{5sec:Governing equations}
We consider a multi-component PIP having inertial positively charged warm ions, (mass $m_1$; charge
$eZ_1$; temperature $T_1$; number density $\tilde{n_1}$), negatively
charged warm ions (mass $m_2$; charge $-eZ_2$; temperature $T_2$; number density $\tilde{n_2}$),
inertialess electrons, featuring non-thermal distribution (mass $m_e$; charge $-e$; temperature $T_e$;
number density $\tilde{n_e}$), inertialess positrons, obeying non-thermal distribution (mass $m_p$;
charge $e$; temperature $T_p$; number density $\tilde{n_p}$)
and static negatively charged massive dust grains (charge $-eZ_d$; number density $n_{d}$);
where $Z_1$ ($Z_2$) is the charge state of the positive (negative) ion, and $Z_d$ is the charge state of
the negative dust grains, and $e$ is the magnitude of the charge of an electron. An external magnetic
field $\mathbf{B}_0$ has  been considered in the system directed along the $z$-axis defining $\mathbf{B}_0 = B_0\hat{z}$,
where $B_0$ and $\hat{z}$ denoted the strength of the external magnetic field and unit vector
directed along the $z$-axis, respectively. The dynamics of the PIP system is governed by the following
set of equations \cite{Adhikary2012,Sahu2014,Atteyaa2018,C2}
\begin{eqnarray}
&&\hspace*{-1.3cm}\frac{\partial \tilde{n}_1}{\partial \tilde{t}}+\acute{\nabla}\cdot(\tilde{n}_1 \tilde{u}_1)=0,
\label{5eq:1}\\
&&\hspace*{-1.3cm}\frac{\partial \tilde{u}_1}{\partial\tilde{t}}+(\tilde{u}_1\cdot\acute{\nabla})\tilde{u}_1=-\frac{Z_1e}{m_1}\acute{\nabla}\tilde{\psi} +\frac{Z_1eB_0}{m_1}(\tilde{u}_1\times\hat{z})
\nonumber\\
&&\hspace*{2.5cm}-\frac{1}{m_1\tilde{n}_1}\acute{\nabla} P_1 +\tilde{\eta}_1\acute{\nabla}^2\tilde{u}_1,
\label{5eq:2}\\
&&\hspace*{-1.3cm}\frac{\partial\tilde{n}_2}{\partial \tilde{t}}+\acute{\nabla}\cdot(\tilde{n}_2\tilde{u}_2)=0,
\label{5eq:3}\\
&&\hspace*{-1.3cm}\frac{\partial\tilde{u}_2}{\partial \tilde{t}}+(\tilde{u}_2\cdot\acute{\nabla})\tilde{u}_{2}=\frac{Z_2e}{m_2}\acute{\nabla}\tilde{\psi} -\frac{Z_2eB_0}{m_2}(\tilde{u}_2\times\hat{z})
\nonumber\\
&&\hspace*{2.5cm}-\frac{1}{m_2\tilde{n}_2}\acute{\nabla} P_2 +\tilde{\eta}_2\acute{\nabla}^2\tilde{u}_2,
\label{5eq:4}\\
&&\hspace*{-1.3cm}\acute{\nabla}^2\tilde{\psi}=4\pi e[\tilde{n}_e+Z_d\tilde{n}_d+Z_2\tilde{n}_2-Z_1\tilde{n}_1-\tilde{n}_p],
\label{5eq:5}\
\end{eqnarray}
where $\tilde{u}_2$ ($\tilde{u}_2$) is the positive (negative) ion fluid velocity;
$\tilde{\eta}_1 = \mu_1/m_1n_1$ ($\tilde{\eta}_2 = \mu_2/m_2n_2$) is the kinematic viscosity of the
positive (negative) ion; $P_1$ ($P_2$) is the pressure of positive (negative) ion, and $\tilde{\psi}$ represents the
electrostatic wave potential. Now, we  are introducing normalized variables, namely, $n_1\rightarrow\tilde{n}_1/n_{10}$, $n_2\rightarrow\tilde{n}_2/n_{20}$, $n_e\rightarrow\tilde{n}_e/n_{e0}$, $n_p\rightarrow\tilde{n}_p/n_{p0}$, $n_d\rightarrow\tilde{n}_d/n_{d0}$, $u_1\rightarrow\tilde{u}_1/C_2$, $u_2\rightarrow\tilde{u}_2/C_2$ [where $C_2=(Z_2k_BT_e/m_2)^{1/2}$, $k_B$ being the Boltzmann constant]; $\psi\rightarrow\tilde{\psi}e/k_BT_e$; $t=\tilde{t}/\omega_{P_2}^{-1}$ [where $\omega_{P_2}^{-1}=(m_2/4\pi e^{2}Z_2^{2}n_{20})^{1/2}$]; $\nabla=\acute{\nabla}/\lambda_{D}$ [where $\lambda_{D}=(k_BT_e/4\pi e^2Z_2n_{20})^{1/2}$]. The pressure term of the positive ion can be recognized as $P_1=P_{10}(N_1/n_{10})^\gamma$ with $P_{10}=n_{10}k_BT_1$ being the equilibrium
pressure of the positive ion, and the pressure term of the negative ion can be recognized as $P_2=P_{20}(N_2/n_{20})^\gamma$ with $P_{20}=n_{20}k_BT_2$ being the equilibrium pressure of the negative ion, and $\gamma=(N+2)/N$ (where $N$ is the degree of freedom and for three-dimensional case
$N=3$, then $\gamma=5/3$). For simplicity, we have considered ($\tilde{\eta}_1\approx\tilde{\eta}_2=\eta$), and $\eta$ is
normalized by $\omega_{p_2}\lambda_D^{2}$. The quasi-neutrality condition at equilibrium for our plasma model
can be written as $Z_1n_{10}+n_{p0}\approx Z_2n_{20}+Z_d n_{d0}+n_{e0}$. Equations \eqref{5eq:1}$-$\eqref{5eq:5} can be
expressed in the normalized form as \cite{El-Labany2020}:
\begin{eqnarray}
&&\hspace*{-1.3cm}\frac{\partial n_1}{\partial t}+\nabla\cdot(n_1u_1)=0,
\label{5eq:6}\\
&&\hspace*{-1.3cm}\frac{\partial u_1}{\partial t}+(u_1\cdot\nabla)u_1=-\alpha_1\nabla\psi+\alpha_1\Omega_c(u_1\times\hat{z})
\nonumber\\
&&\hspace*{2.5cm}-\alpha_2\nabla n_1^{\gamma-1}+\eta\nabla^2u_1,
\label{5eq:7}\\
&&\hspace*{-1.3cm}\frac{\partial n_2}{\partial t}+\nabla\cdot(n_2u_2)=0,
\label{5eq:8}\\
&&\hspace*{-1.3cm}\frac{\partial u_2}{\partial t}+(u_2\cdot\nabla)u_2=\nabla\psi-\Omega_c(u_2\times\hat{z})
\nonumber\\
&&\hspace*{2.5cm}-\alpha_3\nabla n_2^{\gamma-1}+\eta\nabla^2u_2,
\label{5eq:9}\\
&&\hspace*{-1.3cm}\nabla^2\psi=\lambda_en_e-\lambda_pn_p+\lambda_dn_d-(1+\lambda_e+\lambda_d-\lambda_p)n_1+n_2.
\label{5eq:10}\
\end{eqnarray}
Other plasma parameters can be recognized as $\alpha_1=Z_1m_2/Z_2m_1$,
$\alpha_2=5T_1m_2/2Z_2T_em_1$,
$\alpha_3=5T_2/2Z_2T_e$, $\lambda_e = n_{e0}/Z_2n_{20}$, $\lambda_d = Z_d n_{d0}/Z_2n_{20}$, $\lambda_p = n_{p0}/Z_2n_{20}$, and $\Omega_c=\omega_c/\omega_{p_2}$ [where $\omega_c=Z_2eB_0/m_2$].
Now, the expression for the number density of electrons and positrons following non-thermal
distribution can be, respectively, written as  \cite{Cairns1995}
\begin{eqnarray}
&&\hspace*{-1.3cm}n_e=(1-\beta \psi+\beta \psi^2)\mbox{exp}(\psi),
\label{5eq:11}\\
&&\hspace*{-1.3cm}n_p=(1+\beta \alpha_4 \psi+\beta \alpha_4^2 \psi^2)\mbox{exp}(-\alpha_4 \psi),
\label{5eq:12}\
\end{eqnarray}
where $\beta=4\alpha/(1+3\alpha)$, ($\alpha$ represents the number of non-thermal populations in our considered model) and $\alpha_4=T_e/T_p$.
Now, by substituting Eqs. \eqref{5eq:11}-\eqref{5eq:12} into the Eq. \eqref{5eq:10}, and expanding up to third order in $\psi$, we can write
\begin{eqnarray}
&&\hspace*{-1.3cm}\nabla^2 \psi=\lambda_e-\lambda_p+n_d\lambda_d+n_2-\Lambda n_1+\sigma_1\psi+\sigma_2\psi^2+\cdot\cdot\cdot,
\label{5eq:13}\
\end{eqnarray}
where $\Lambda=1+\lambda_e+\lambda_d-\lambda_p$, $\sigma_1=\lambda_e(1-\beta)-\lambda_p \alpha_4 (\beta-1)$,
and $\sigma_2=\lambda_e/2-\lambda_p \alpha_4^2/2$. We note that the terms containing $\sigma_1$ and $\sigma_2$ are the contribution of
non-thermal distributed electrons and positrons.
\section{Derivation of the Burgers' equation}
\label{5sec:Derivation of the Burgers equation}
To derive the Burgers' equation for the DIASHWs propagating in a PIP,
we first introduce the stretched co-ordinates \cite{Washimi1966,Hossen2017aa} as
\begin{eqnarray}
&&\hspace*{-1.3cm}\xi=\epsilon(l_xx+l_yy+l_zz-V_p t),
\label{5eq:14}\\
&&\hspace*{-1.3cm}\tau={\epsilon}^2 t,
\label{5eq:15}\
\end{eqnarray}
where $V_p$ is the phase speed and $\epsilon$ is a smallness parameter denoting the weakness of
the dissipation ($0<\epsilon<1$). It is noted that $l_x$, $l_y$, and $l_z$ (i.e., $l_x^2+l_y^2+l_z^2=1$) are
the directional cosines of the wave vector of $k$ along $x$, $y$, and $z$-axes, respectively. Then,
the dependent variables can be expressed in power series of $\epsilon$ as \cite{Hossen2017aa}
\begin{eqnarray}
&&\hspace*{-1.3cm}n_1=1+\epsilon n_1^{(1)}+\epsilon^2 n_1^{(2)}+\epsilon^3 n_1^{(3)}+\cdot\cdot\cdot,
\label{5eq:16}\\
&&\hspace*{-1.3cm}n_2=1+\epsilon n_2^{(1)}+\epsilon^2 n_2^{(2)}+\epsilon^3 n_2^{(3)}+\cdot\cdot\cdot,
\label{5eq:17}\\
&&\hspace*{-1.3cm}u_{1x,y}=\epsilon^2 u_{1x,y}^{(1)}+\epsilon^3 u_{1x,y}^{(2)}+\cdot\cdot\cdot,
\label{5eq:18}\\
&&\hspace*{-1.3cm}u_{2x,y}=\epsilon^2 u_{2x,y}^{(1)}+\epsilon^3 u_{2x,y}^{(2)}+\cdot\cdot\cdot,
\label{5eq:19}\\
&&\hspace*{-1.3cm}u_{1z}=\epsilon u_{1z}^{(1)}+\epsilon^2 u_{1z}^{(2)}+\cdot\cdot\cdot,
\label{5eq:20}\\
&&\hspace*{-1.3cm}u_{2z}=\epsilon u_{2z}^{(1)}+\epsilon^2 u_{2z}^{(2)}+\cdot\cdot\cdot,
\label{5eq:21}\\
&&\hspace*{-1.3cm}\psi=\epsilon \psi^{(1)}+\epsilon^2\psi^{(2)}+\cdot\cdot\cdot.
\label{5eq:22}\
\end{eqnarray}
Now, by substituting Eqs. \eqref{5eq:14}$-$\eqref{5eq:22} into Eqs. \eqref{5eq:6}$-$\eqref{5eq:9}, and
\eqref{5eq:13}, and collecting the terms containing $\epsilon$, the first-order equations become
\begin{eqnarray}
&&\hspace*{-1.3cm} n_1^{(1)}=\frac{3\alpha_1l_z^2}{3V_p^2-2\alpha_2l_z^2}\psi^{(1)},
\label{5eq:23}\\
&&\hspace*{-1.3cm}u_{1z}^{(1)}=\frac{3V_p\alpha_1l_z}{3V_p^2-2\alpha_2l_z^2}\psi^{(1)},
\label{5eq:24}\\
&&\hspace*{-1.3cm}n_2^{(1)}=-\frac{3l_z^2}{3V_p^2-2\alpha_3l_z^2}\psi^{(1)},
\label{5eq:25}\\
&&\hspace*{-1.3cm}u_{2z}^{(1)}=-\frac{3V_pl_z}{3V_p^2-2\alpha_3l_z^2}\psi^{(1)}.
\label{5eq:26}\
\end{eqnarray}
Now, the phase speed of DIASHWs can be read as
\begin{eqnarray}
&&\hspace*{-1.3cm}V_{p}\equiv V_{p+}=l_z\sqrt{{\frac{-a_1+\sqrt{a_1^2-36\sigma_1a_2}}{18\sigma_1}}},
\label{5eq:27}\\
&&\hspace*{-1.3cm}V_{p}\equiv V_{p-}=l_z\sqrt{{\frac{-a_1-\sqrt{a_1^2-36\sigma_1a_2}}{18\sigma_1}}},
\label{5eq:28}\
\end{eqnarray}
where $a_1=-{6\sigma_1\alpha_2+6\sigma_1\alpha_3+9+\Lambda n_1}$ and $a_2=4\sigma_1\alpha_2\alpha_3+6\alpha_2+6\alpha_1\alpha_3\Lambda$.
The $x$ and $y$-components of the first-order momentum equations can be written as
\begin{eqnarray}
&&\hspace*{-1.3cm}u_{1x}^{(1)}=-\frac{3l_yV_p^2}{\Omega_{c}(3V_p^2-2\alpha_2l_z^2)}~\frac{\partial\psi^{(1)}}{\partial\xi},
\label{5eq:29}\\
&&\hspace*{-1.3cm}u_{1y}^{(1)}=\frac{3l_xV_p^2}{\Omega_{c}(3V_p^2-2\alpha_2l_z^2)}~\frac{\partial\psi^{(1)}}{\partial\xi},
\label{5eq:30}\\
&&\hspace*{-1.3cm}u_{2x}^{(1)}=-\frac{3l_yV_p^2}{\Omega_{c}(3V_p^2-2\alpha_3l_z^2)}~\frac{\partial\psi^{(1)}}{\partial\xi},
\label{5eq:31}\\
&&\hspace*{-1.3cm} u_{2y}^{(1)}=\frac{3l_xV_p^2}{\Omega_{c}(3V_p^2-2\sigma_3l_z^2)}~\frac{\partial \psi^{(1)}}{\partial\xi}.
\label{5eq:32}\
\end{eqnarray}
Now, by following the next higher-order terms, the equation of continuity, momentum equation, and Poisson's equation can be written as
\begin{eqnarray}
&&\hspace*{-1.3cm}\frac{\partial n_1^{(1)}}{\partial\tau}-V_p\frac{\partial n_1^{(2)}}{\partial\xi}+l_x\frac{\partial u_{1x}^{(1)}}{\partial\xi}+l_y\frac{\partial u_{1y}^{(1)}}{\partial\xi}
\nonumber\\
&&\hspace*{1.5cm}+l_z\frac{\partial u_{1z}^{(2)}}{\partial\xi}+l_z\frac{\partial}{\partial\xi}\big(n_1^{(1)}u_{+z}^{(1)}\big)=0,
\label{5eq:33}\\
&&\hspace*{-1.3cm}\frac{\partial u_{1z}^{(1)}}{\partial\tau}-V_p\frac{\partial u_{1z}^{(2)}}{\partial\xi}+l_zu_{1z}^{(1)}\frac{\partial u_{1z}^{(1)}}{\partial\xi}+\alpha_1l_z\frac{\partial\psi^{(2)}}{\partial\xi}
\nonumber\\
&&\hspace*{0.5cm}+\alpha_2l_z\frac{\partial }{\partial\xi}\bigg[\frac{2}{3}n_1^{(2)}-\frac{1}{9}(n_1^{(1)})^2\bigg]-\eta\frac{\partial^2u_{1z}^{(1)}}{\partial\xi^2}=0,
\label{5eq:34}\\
&&\hspace*{-1.3cm}\frac{\partial n_2^{(1)}}{\partial\tau}-V_p\frac{\partial n_2^{(2)}}{\partial\xi}+l_x\frac{\partial u_{2x}^{(1)}}{\partial\xi}+l_y\frac{\partial u_{2y}^{(1)}}{\partial\xi}
\nonumber\\
&&\hspace*{1.5cm}+l_z\frac{\partial u_{2z}^{(2)}}{\partial\xi}+l_z\frac{\partial}{\partial\xi}\big(n_2^{(1)}u_{2z}^{(1)}\big)=0,
\label{5eq:35}\\
&&\hspace*{-1.3cm}\frac{\partial u_{2z}^{(1)}}{\partial\tau}-V_p\frac{\partial u_{2z}^{(2)}}{\partial\xi}+l_zu_{2z}^{(1)}\frac{\partial u_{2z}^{(1)}}{\partial\xi}-l_z\frac{\partial\psi^{(2)}}{\partial\xi}
\nonumber\\
&&\hspace*{0.5cm}+\alpha_3l_z\frac{\partial }{\partial\xi}\bigg[\frac{2}{3}n_2^{(2)}-\frac{1}{9}(n_2^{(1)})^2\bigg]-\eta\frac{\partial^2u_{2z}^{(1)}}{\partial\xi^2}=0,
\label{5eq:36}\\
&&\hspace*{-1.3cm}\sigma_1\psi^{(2)}+\sigma_2{[\psi^{(1)}]}^2+n_2^{(2)}-\Lambda n_1^{(2)}=0.
\label{5eq:37}\
\end{eqnarray}
Finally, the next higher-order terms of Eqs. \eqref{5eq:6}$-$\eqref{5eq:9}, and \eqref{5eq:12}, with the help of
Eqs. \eqref{5eq:23}$-$\eqref{5eq:37}, can provide the Burgers' equation as
\begin{eqnarray}
&&\hspace*{-1.3cm} \frac{\partial\Psi}{\partial\tau}+A\Psi\frac{\partial\Psi}{\partial\xi}=C\frac{\partial^2\Psi}{\partial\xi^2},
\label{5eq:38}\
\end{eqnarray}
where $\Psi=\psi^{(1)}$ is used for simplicity. In Eq. \eqref{5eq:38}, the nonlinear coefficient ($A$) and dissipative coefficient ($C$) are given by the following expression
\begin{eqnarray}
&&\hspace*{-1.3cm}A=\frac{(M_1S_1^3-M_2S_2^3-2\sigma_2S_1^3S_2^3)}{M_3S_1S_2},~~\mbox{and}~~C =\frac{\eta}{2},
\label{5eq:39}\
\end{eqnarray}
where $M_1 =\Lambda(81\alpha_1^2V_p^2l_z^4-6\alpha_2\alpha_1^2l_z^6)$, $M_2=81V_p^2l_z^4-6\alpha_3l_z^6$, and $M_3=18V_pl_z^2{1+\alpha_1\Lambda}$.
Now, we look forward to the stationary shock wave solution of this Burgers' equation by
taking $\zeta =\xi-U_0\tau'$ and $\tau =\tau'$, where $U_0$ is the speed of the shock waves in the reference frame.
These allow us to represent the stationary shock wave solution as \cite{Karpman1975,Hasegawa1975,Hossen2017aa}
\begin{eqnarray}
&&\hspace*{-1.3cm}\Psi=\Psi_m \Big[1-\tanh\bigg(\frac{\zeta}{\Delta}\bigg)\Big],
\label{5eq:40}\
\end{eqnarray}
where $\Psi_m$ is the amplitude and $\Delta$ is the width. The expression of the amplitude and width can be given by the following equations
\begin{eqnarray}
&&\hspace*{-1.3cm}\Psi_m=\frac{U_0}{A},~~~~\mbox{and}~~~~\Delta=\frac{2C}{U_0}.
\label{5eq:41}\
\end{eqnarray}
\begin{figure}
\centering
\includegraphics[width=80mm]{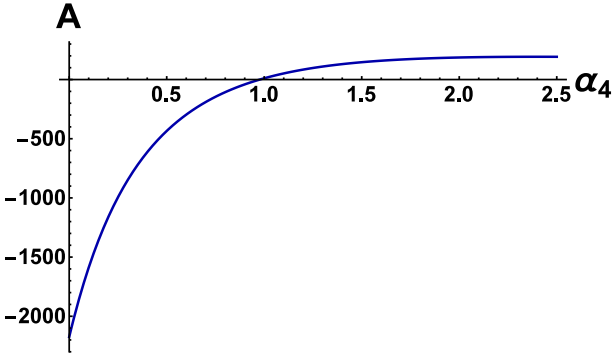}
\caption{The variation of nonlinear coefficient $A$ with $\alpha_4$ along with $\alpha_1=1.5$,
$\alpha_2 =0.05$, $\alpha_3=0.03$, $\lambda_p=1.5$, $\lambda_e=1.7$, $\lambda_d=0.05$, $\delta = 30^\circ$, $\alpha=0.5$, and $V_{p}\equiv V_{p+}$.}
\label{5Fig:F1}
\end{figure}
\begin{figure}
\centering
\includegraphics[width=80mm]{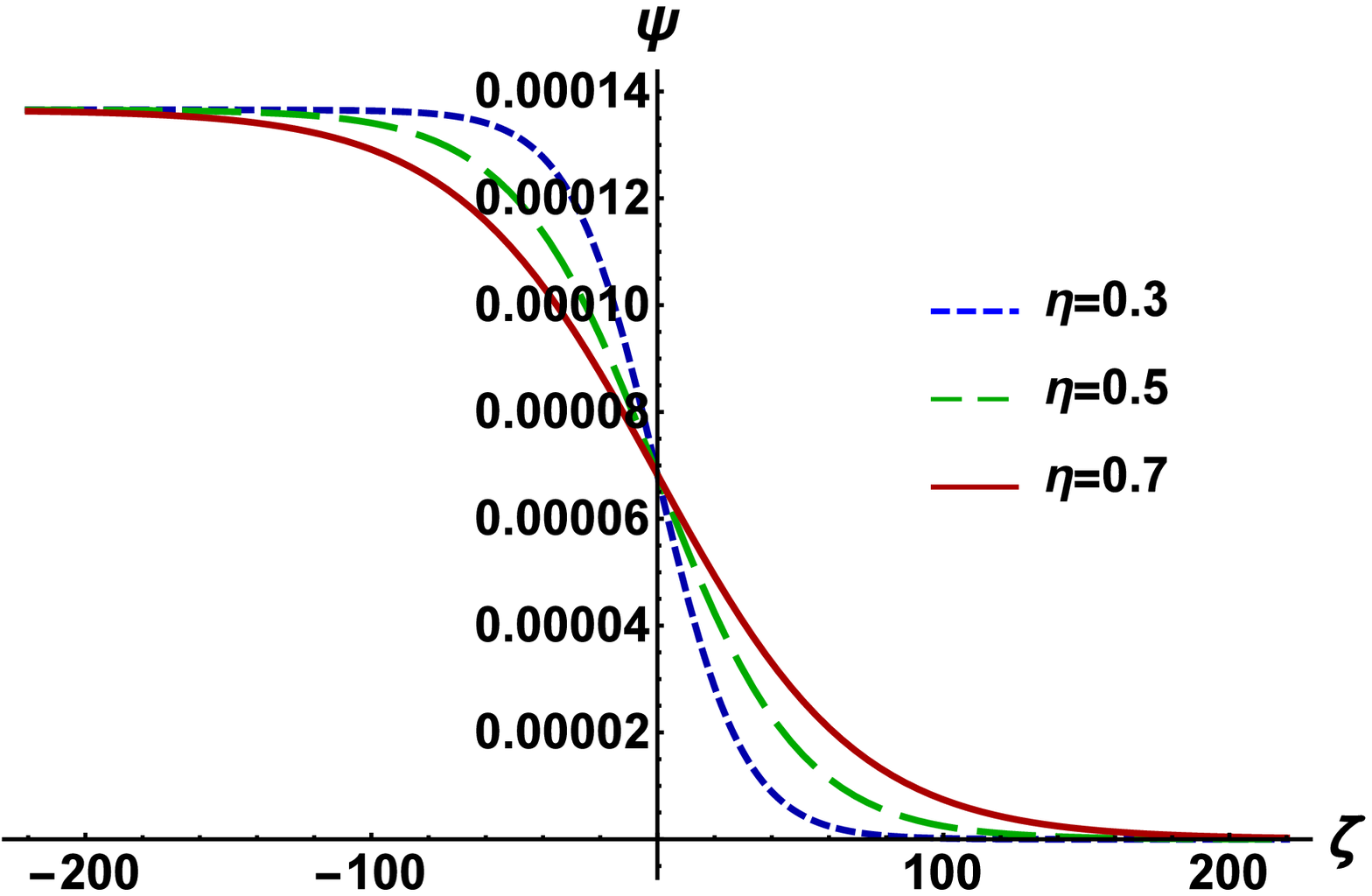}
\caption{The variation of $\Psi$ with $\zeta$ for different values of $\eta$  under consideration of $\alpha_4>\alpha_{4c}$
along with $\alpha_1=1.5$, $\alpha_2 =0.05$, $\alpha_3=0.03$, $\alpha_4=1.5$, $\lambda_p=1.5$,
$\lambda_e=1.7$, $\lambda_d=0.05$, $\delta = 30^\circ$, $\alpha=0.5$,  $U_0=0.01$, and $V_{p}\equiv V_{p+}$.}
\label{5Fig:F2}
\end{figure}
\begin{figure}
\centering
\includegraphics[width=80mm]{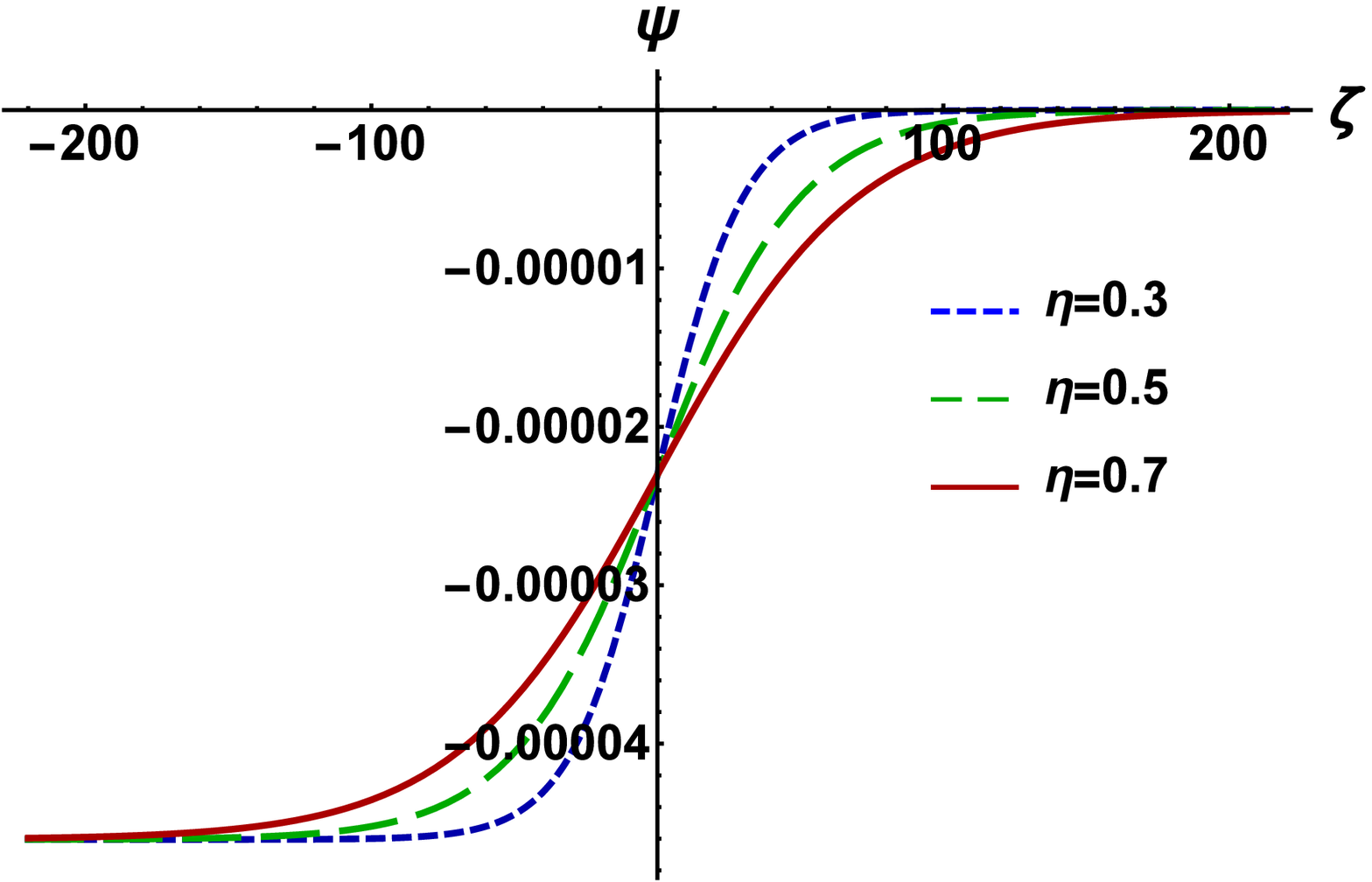}
\caption{The variation of $\Psi$ with $\zeta$ for different values of $\eta$  under consideration of $\alpha_4<\alpha_{4c}$
along with $\alpha_1=1.5$, $\alpha_2 =0.05$, $\alpha_3=0.03$, $\alpha_4=0.5$, $\lambda_p=1.5$,
$\lambda_e=1.7$, $\lambda_d=0.05$, $\delta = 30^\circ$, $\alpha=0.5$,  $U_0=0.01$, and $V_{p}\equiv V_{p+}$.}
\label{5Fig:F3}
\end{figure}
\begin{figure}
\centering
\includegraphics[width=80mm]{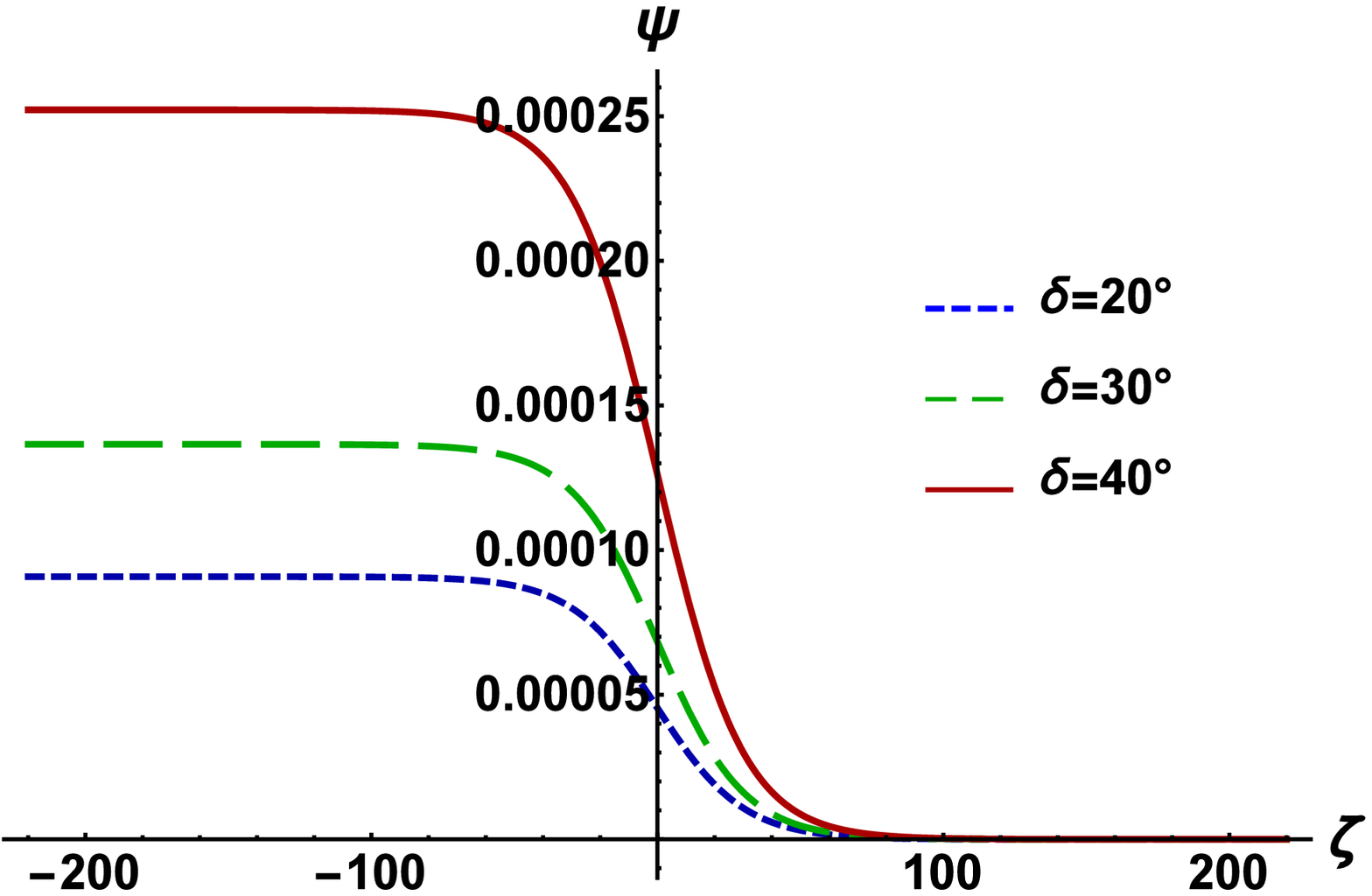}
\caption{The variation of $\Psi$ with $\zeta$ for different values of $\delta$ under consideration of $\alpha_4>\alpha_{4c}$ along with $\alpha_1=1.5$, $\alpha_2 =0.05$, $\alpha_3=0.03$, $\alpha_4=1.5$, $\lambda_p=1.5$,
$\lambda_e=1.7$, $\lambda_d=0.05$, $\eta=0.3$, $\alpha=0.5$, $U_0=0.01$, and $V_{p}\equiv V_{p+}$.}
\label{5Fig:F4}
\end{figure}
\begin{figure}
\centering
\includegraphics[width=80mm]{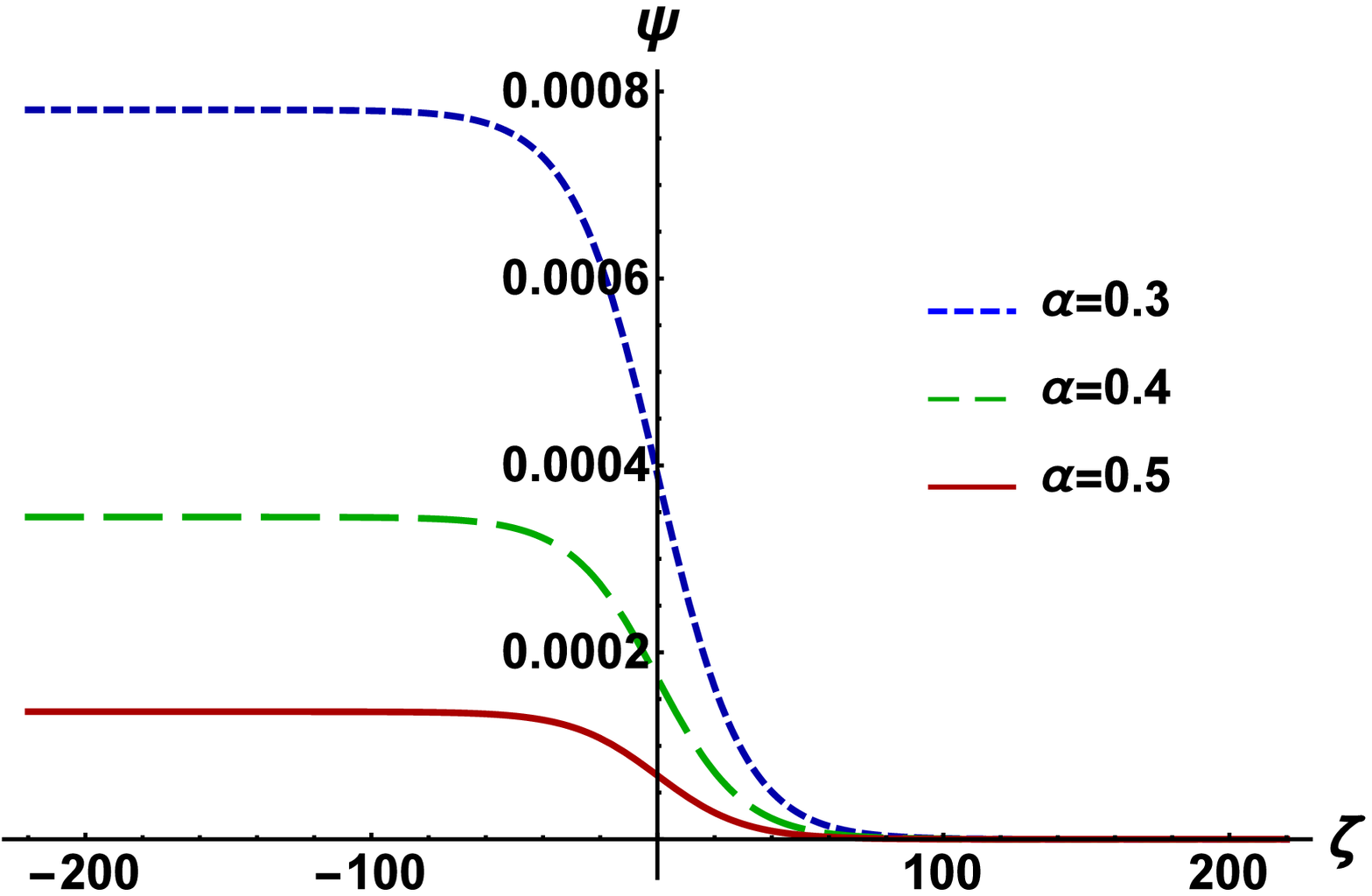}
\caption{The variation of $\Psi$ with $\zeta$ for different values of $\alpha$  under consideration of $\alpha_4>\alpha_{4c}$
along with $\alpha_1=1.5$, $\alpha_2 =0.05$, $\alpha_3=0.03$, $\alpha_4=1.5$, $\lambda_p=1.5$,
$\lambda_e=1.7$, $\lambda_d=0.05$, $\delta = 30^\circ$, $\eta=0.3$, $U_0=0.01$, and $V_{p}\equiv V_{p+}$.}
\label{5Fig:F5}
\end{figure}
\begin{figure}
\centering
\includegraphics[width=80mm]{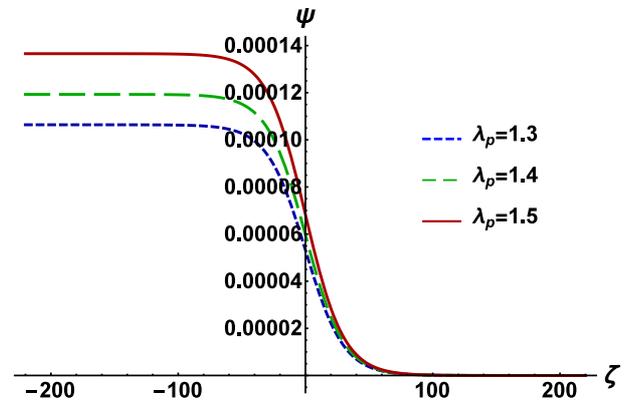}
\caption{The variation of $\Psi$ with $\zeta$ for different values of $\lambda_p$  under consideration of $\alpha_4>\alpha_{4c}$
along with $\alpha_1=1.5$, $\alpha_2 =0.05$, $\alpha_3=0.03$, $\alpha_4=1.5$, $\lambda_e=1.7$, $\lambda_d=0.05$, $\delta = 30^\circ$, $\eta=0.3$,  $\alpha=0.5$, $U_0=0.01$, and $V_{p}\equiv V_{p+}$.}
\label{5Fig:F6}
\end{figure}
\section{Numerical Analysis and Discussion}
\label{5sec:Numerical Analysis and Discussion}
Now, we would like to observe the basic properties of DIASHWs in a magnetized PIP having inertial pair-ions, inertialess non-thermal
distributed electrons and positrons, and static negatively charged massive dust grains by changing the various plasma parameters,
viz., ion kinematic  viscosity, oblique angle, non-thermality of electrons and positrons, mass, charge, temperature, and number density
of the plasma species.  Equation \eqref{5eq:41} shows that under consideration $U_0>0$ and $C>0$, no shock wave will exist if $A=0$
as the amplitude of the wave becomes infinite which clearly violates the reductive perturbation method. So, $A$ can be positive (i.e., $A>0$)
or negative (i.e., $A<0$) according to the value of other plasma parameters. Figure \ref{5Fig:F1} illustrates the variation of
$A$ with $\alpha_4$, and it is obvious from this figure that $A$ can be negative, zero, and  positive according to the values of $\alpha_4$
when other plasma parameters are $\alpha_1=1.5$, $\alpha_2 =0.05$, $\alpha_3=0.03$, $\lambda_p=1.5$, $\lambda_e=1.7$, $\lambda_d=0.05$,
$\delta=30^\circ$, and $\alpha=0.5$. The point at which $A$ becomes zero for the value of $\alpha_4$ is known as the critical value of
$\alpha_4$ (i.e., $\alpha_{4c}$). In our present analysis, the critical value of $\alpha_4$ is $\alpha_{4c}\equiv 01$. So, the negative (positive) shock profile can be exist for the value of $\alpha_{4}<\alpha_{4c}$ ($\alpha_{4}>\alpha_{4c}$).

\begin{figure}
\centering
\includegraphics[width=80mm]{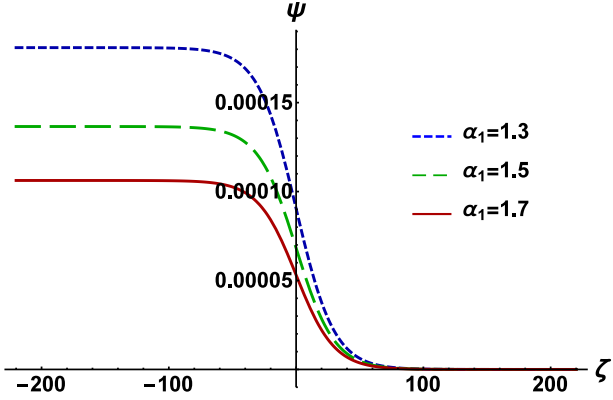}
\caption{The variation of $\Psi$ with $\zeta$ for different values of $\alpha_1$  under consideration of $\alpha_4>\alpha_{4c}$
along with $\alpha_2 =0.05$, $\alpha_3=0.03$, $\alpha_4=1.5$, $\lambda_e=1.7$, $\lambda_p=1.5$, $\lambda_d=0.05$, $\delta = 30^\circ$, $\eta=0.3$,  $\alpha=0.5$, $U_0=0.01$, and $V_{p}\equiv V_{p+}$.}
\label{5Fig:F7}
\end{figure}
Figures \ref{5Fig:F2} and \ref{5Fig:F3} respectively represent the variation of the positive and negative shock profiles with ion kinematic
viscosity (via $\eta$) when other plasma parameters are remain constant. It is really interestingly that the steepness of both
positive and negative shock profiles declines with the increase of $\eta$ without affecting the height.
Figure \ref{5Fig:F4} describes the effects of the external magnetic field to the formation of the positive shock profile.
The increase in oblique angle rises the height of the positive shock profile and this result is analogous to the result of Ref. \cite{Shahmansouri2014}.

The height of the positive shock profile is so much sensitive to the change of non-thermality of the electrons and positrons which can be
seen in Fig. \ref{5Fig:F5}. There is a decrease in the amplitude of positive shock profile when electrons and positrons deviate from thermodynamic equilibrium, and this result is compatible with the result of Ref. \cite{Alinejad2010}. The variation of the DIASHWs with negative ion charge state, negative ion and positron number densities (via $\lambda_p$) can be observed in Fig. \ref{5Fig:F6}. It is clear from Fig. \ref{5Fig:F6} that as we increase the positron (negative ion) number density, the height of the positive shock wave increases (decreases) when the charge of the negative ion remains constant or the the height of the positive shock wave  decreases with the charge of the negative ion for a fixed value of the number density of
positron and negative ion.

\begin{figure}
\centering
\includegraphics[width=80mm]{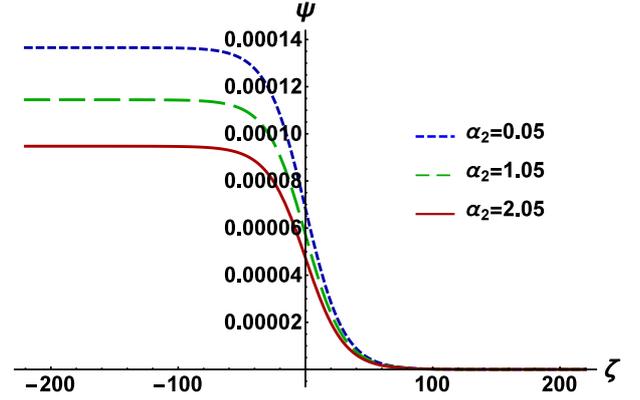}
\caption{The variation of $\Psi$ with $\zeta$ for different values of $\alpha_2$  under consideration of $\alpha_4>\alpha_{4c}$
along with $\alpha_1 =1.5$, $\alpha_3=0.03$, $\alpha_4=1.5$, $\lambda_e=1.7$, $\lambda_p=1.5$, $\lambda_d=0.05$, $\delta = 30^\circ$, $\eta=0.3$,  $\alpha=0.5$, $U_0=0.01$, and $V_{p}\equiv V_{p+}$.}
\label{5Fig:F8}
\end{figure}
The charge and mass of the positive and negative ions are rigourously responsible to change the height of the positive shock profile.
The variation of the DIASHWs with $\alpha_1$ has been demonstrated in Fig. \ref{5Fig:F7}, and it is obvious from this figure that
the height of the positive shock profile increases (decreases) with increasing the value of positive (negative) ion mass for a fixed value of the
their charge state. But as we increase the charge state of the negative (positive) ion then the amplitude of the positive shock profile
increases (decreases) when their mass are constant. The effects of the temperature of electron and positive ion (via $\alpha_2$) can be seen in Fig. \ref{5Fig:F8}, and the amplitude of the shock profile enhances (diminishes) with electron (positive ion) temperature when other plasma parameters are invariant.
\section{Conclusion}
\label{5sec:Conclusion}
In our present investigation, we have considered a multi-component magnetized PIP having static dust grains, non-thermal electrons and positrons.
The Burgers' equation has been derived by employing  reductive perturbation method \cite{C3} for studying DIASHWs. The results that we have found from this investigation can be summarized as follows:
\begin{itemize}
\item The negative (positive) shock profile can be exist for the value of $\alpha_{4}<\alpha_{4c}$ ($\alpha_{4}>\alpha_{4c}$).
\item The steepness of both positive and negative shock profiles declines with the increase of $\eta$ without affecting the height.
\item The increase in oblique angle rises the height of the positive shock profile.
\item The height of the positive shock wave increases with the number density of positron.
\item The temperature of the electrons enhances the amplitude of the shock profile.
\end{itemize}
The results are applicable in understanding the criteria for the formation of DIASHWs in astrophysical plasmas,
viz., cometary comae \cite{Chaizy1991}, upper regions of Titan's atmosphere \cite{Coates2007,Massey1976,Sabry2009},
plasmas in the D and F-regions of Earth's ionosphere \cite{Massey1976,Sabry2009,Abdelwahed2016}, and also in
laboratory environments, viz., ($K^+$, $SF_6^-$) plasma \cite{Song1991,Sato1994}, ($Ar^+$, $F^-$) plasma \cite{Nakamura1984}, plasma processing reactors \cite{Gottscho1986}, plasma etching \cite{Sheehan1988}, combustion products \cite{Sheehan1988}, ($Xe^+$, $F^-$) plasma \cite{Ichiki2002}, neutral beam sources \cite{Bacal1979}, ($Ar^+$, $SF_6^-$) plasma \cite{Wong1975,Cooney1991,Nakamura1997},($Ar^+$, $O_2^-$) plasma, and Fullerene ($C_{60}^+$, $C_{60}^-$) plasma \cite{Oohara2003,Hatakeyama2005}, etc.


\begin{thebibliography}{99}

\bibitem{Shukla2002} P.K. Shukla and A.A. Mamun, \textit{Introdustion to Dusty Plasma Physics} (Institute of Physics, Bristol, 2002).

\bibitem{Chaizy1991}P.H. Chaizy, \textit{et al.}, Nature (London), \textbf{349}, 393 (1991).

\bibitem{Coates2007}A. J. Coates, \textit{et al.}, Geophys. Res. Lett. \textbf{34},
    L22103 (2007).

\bibitem{Massey1976}H. Massey, \textit{Negative Ions}, 3rd ed., (Cambridge University Press, Cambridge, 1976)

\bibitem{Sabry2009} R. Sabry, \textit{et al.}, Phys. Plasmas \textbf{16}, 032302 (2009).

\bibitem{Abdelwahed2016} H.G. Abdelwahed, \textit{et al.}, Phys. Plasmas \textbf{23}, 022102 (2016).

\bibitem{Song1991} B. Song, \textit{et al.}, Phys. Fluids B \textbf{3}, 284 (1991).

\bibitem{Sato1994} N. Sato, Plasma Sources Sci. Technol. \textbf{3}, 395 (1994).

\bibitem{Nakamura1984} Y. Nakamura and I. Tsukabayashi, Phys. Rev. Lett. \textbf{52}, 2356 (1984).

\bibitem{Gottscho1986} R.A. Gottscho and C.E. Gaebe, IEEE Trans. Plasma Sci. \textbf{14}, 92 (1986).

\bibitem{Sheehan1988} D.P. Sheehan and N. Rynn, Rev. Sci. lnstrum. \textbf{59}, 8 (1988).

\bibitem{Ichiki2002} R. Ichiki, \textit{et al.}, Phys. Plasmas \textbf{9}, 4481 (2002).

\bibitem{Bacal1979} M. Bacal and G.W. Hamilton, Phys. Rev. Lett. \textbf{42}, 1538 (1979).

\bibitem{Wong1975} A.Y. Wong, D.L. Mamas, and D. Arnush, Phys. Fluids \textbf{18}, 1489 (1975).

\bibitem{Cooney1991} J.L. Cooney, \textit{et al.}, Phys. Fluids B \textbf{3}, 2758 (1991).

\bibitem{Nakamura1997} Y. Nakamura, \textit{et al.}, Plasma Phys. Control. Fusion \textbf{39}, 105 (1997).

\bibitem{Oohara2003} W. Oohara and R. Hatakeyama, Phys. Rev. Lett. \textbf{91}, 205005 (2003).

\bibitem{Hatakeyama2005} R. Hatakeyama and W. Oohara, Phys. Scripta \textbf{116}, 101 (2005).

\bibitem{Shukla1992}P.K. Shukla and V.P. Silin,  Phys. Scr. \textbf{45}, 508 (1992).

\bibitem{Shukla2000} P.K. Shukla, Phys. Plasmas \textbf{7}, 1044 (2000).

\bibitem{Baluku2010} T.K. Baluku, \textit{et al.}, Phys. Plasmas \textbf{17}, 053702 (2010).

\bibitem{Sultana2014} S. Sultana, \textit{et al.},  Astrophys. space Sci \textbf{351}, 581 (2014).

\bibitem{Yasmin2013}S. Yasmin \textit{et al.}, Astrophys Space Sci \textbf{343}, 245 (2013).

\bibitem{Rahman2015}A. Rahman, \textit{et al.}, IEEE Transactions on Plasma Sci. \textbf{43}, 974 (2015).

\bibitem{Abdelsalam2008}U.M. Abdelsalam, \textit{et al.}, Physics Letters A \textbf{372}, 4057 (2008).

\bibitem{Cairns1995}R.A. Carins, \textit{et al.}, \textit{et al.} Geophys. Res. Lett. \textbf{22}, 2709 (1995).

\bibitem{C1} N.M. Heera, \textit{et al.}, AIP Adv. \textbf{11}, 055117 (2021);
             M.H. Rahman,\textit{et al.}, Phys. Plasmas \textbf{25}, 102118 (2018);
             J. Akter,  \textit{et al.} Dust-acoustic envelope solitons and rogue waves
             in an electron depleted plasma. Indian J. Phys (2021). https://doi.org/10.1007/s12648-020-01927-9;
             N.A. Chowdhury, \textit{et al.}, Phys. plasmas \textbf{24}, 113701 (2017);
             M.H. Rahman, \textit{et al.}, Chin. J. Phys. \textbf{56}, 2061 (2018);
             N.A. Chowdhury, \textit{et al.}, Plasma Phys. Rep. \textbf{45}, 459 (2019);
             S. Jahan, \textit{et al.}, Plasma Phys. Rep. \textbf{46}, 90 (2020).

\bibitem{Haider2019}M.M. Haider, \textit{et al.}, Theoretical Phys. \textbf{4}, 124 (2019).

\bibitem{Pakzad2009}H.R. Pakzad and K. Javidan, Pranama Journal of Phys. \textbf{73}, 913 (2009).

\bibitem{Ghai2018}Y. Ghai, \textit{et al.}, Physics of Plasmas \textbf{25}, 013704 (2018).

\bibitem{Sabetar2015} A. Sabetkar and D. Dorranian, J. Theor. Appl. Phys. \textbf{9}, 150 (2015).

\bibitem{Shahmansouri2014} M. Shahmansouri and A.A. Mamun, J. Plasma Physics \textbf{80}, 593 (2014).

\bibitem{Malik2020} H.K. Malik, \textit{et al.}, J. Taibah Univ. Sci. \textbf{14}, 417 (2020).  

\bibitem{Bedi2010} C. Bedi, \textit{et al.}, J. Phys.: Conf. Ser. \textbf{208}, 012037 (2010).

\bibitem{Adhikary2012} N.C. Adhikary, Phys. Lett. A \textbf{376}, 1460 (2012).

\bibitem{Sahu2014}B. Sahu, \textit{et al.}, Phys. Plasmas \textbf{21}, 103701 (2014).

\bibitem{Atteyaa2018} A. Atteya, \textit{et al.}, Chin. J. Phys. \textbf{56}, 1931 (2018).

\bibitem{C2} T. Yeashna, \textit{et al.}, Eur. Phys. J. D \textbf{75}, 135 (2021);
             B.E. Sharmin, \textit{et al.}, Results Phys. \textbf{26}, 104373 (2021);
             S.K. Paul, \textit{et al.}, Pramana J. Phys. \textbf{94}, 58 (2020);
             N.A. Chowdhury, \textit{et al.}, Chaos \textbf{27}, 093105 (2017);
             N.A. Chowdhury, \textit{et al.}, Contrib. Plasma Phys. \textbf{58}, 870 (2018);
             M. Hassan, \textit{et al.}, Commun. Theor. Phys. \textbf{71}, 1017 (2019);
             D.M.S. Zaman, \textit{et al.}, High Temp. \textbf{58}, 789 (2020).

\bibitem{El-Labany2020}S.K. El-Labany, \textit{et al.}, Eur. Phys. J. D \textbf{74}, 104 (2020).

\bibitem{Washimi1966} H. Washimi, T. Taniuti, Phys. Rev. Lett. \textbf{17}, 996 (1966).

\bibitem{Hossen2017aa}M.M. Hossen, \textit{et al.}, High Energy Density Phys. \textbf{24}, 9 (2017).

\bibitem{Karpman1975} V.I. Karpman, \textit{Nonlinear Waves in Dispersive Media}, (Pergamon Press, Oxford, 1975).

\bibitem{Hasegawa1975} A. Hasegawa, \textit{Plasma Instabilities and Nonlinear Effects}, (Springer-Verlag, Berlin, 1975).

\bibitem{Alinejad2010} H. Alinejad, Astrophys Space Sci \textbf{327}, 131 (2010).

\bibitem{C3} S. Jahan, \textit{et al.}, Commun. Theor. Phys. \textbf{71}, 327 (2019);
             T.I. Rajib, \textit{et al.}, Phys. plasmas \textbf{26}, 123701 (2019);
             N. Ahmed, \textit{et al.}, Chaos \textbf{28}, 123107 (2018);
             N.A. Chowdhury, \textit{et al.}, Vacuum \textbf{147}, 31 (2018);
             R.K. Shikha, \textit{et al.}, Eur. Phys. J. D \textbf{73}, 177 (2019);
             S. Banik, \textit{et al.}, Eur. Phys. J. D \textbf{75}, 43 (2021).

\end{thebibliography}
\end{document}